\documentclass[12pt]{article}

\usepackage[mathscr]{eucal}
\usepackage{epsfig,amsfonts}
\usepackage[fleqn]{amsmath}
\usepackage{amsthm,amssymb}
\usepackage{graphicx}
\usepackage{hhline}
\usepackage{cite}
\usepackage{psfrag}

\makeatletter
\@addtoreset{equation}{section}
\makeatother

\def\be{\begin{equation}}
\def\ee{\end{equation}}
\def\bdm{\begin{displaymath}}
\def\edm{\end{displaymath}}
\def\bea{\begin{eqnarray}}
\def\eea{\end{eqnarray}}

\def\ri{{\rm i}}

\def\XXint#1#2#3{{\setbox0=\hbox{$#1{#2#3}{\int}$}
    \vcenter{\hbox{$#2#3$}}\kern-.5\wd0}}

\newcommand{\rd}{\mbox{d}}
\newcommand{\re}{\mbox{e}}

\begin{document}

\begin{titlepage}
\begin{flushright}
RUNHETC-2014-09 \\
\end{flushright}

\vspace{1.cm}

\begin{center}
\begin{LARGE}



{\bf Winding vacuum energies in  \\
a deformed $O(4)$ sigma model}

\end{LARGE}
\vspace{1.3cm}

\begin{large}

{\bf Vladimir V. Bazhanov$^{1,2}$, Gleb A. Kotousov$^{1}$
\\
and 
\\ Sergei  L. Lukyanov}$^{3,4}$

\end{large}

\vspace{1.cm}
$^1$Department of Theoretical Physics,\\
         Research School of Physics and Engineering,\\
    Australian National University, Canberra, ACT 0200, Australia\\\ \\
$^2$Mathematical Sciences Institute,\\
      Australian National University, Canberra, ACT 0200,
      Australia\\\ \\
${}^{3}$NHETC, Department of Physics and Astronomy\\
     Rutgers University\\
     Piscataway, NJ 08855-0849, USA\\
\vspace{.2cm}
and\\
\vspace{.2cm}
${}^{4}$L.D. Landau Institute for Theoretical Physics\\
  Chernogolovka, 142432, Russia\\
\vspace{1.0cm}

\end{center}

\vspace{0.7cm}

\begin{center}
\centerline{\bf Abstract} \vspace{.8cm}
\parbox{13cm}{
We consider the problem of calculating the Casimir energies in the 
winding sectors of  Fateev's $SS$-model, 
which is an integrable two-parameter deformation of the $O(4)$ non-linear  
sigma model in two dimensions. 
This problem lies beyond the scope of all traditional methods of integrable quantum field theory
including the thermodynamic Bethe ansatz and non-linear integral equations.
Here we propose a solution based on a remarkable correspondence between 
classical and quantum integrable systems and express 
the winding energies in terms of certain  solutions of 
the classical sinh-Gordon equation. }
\end{center}

\vfill

\end{titlepage}
\newpage


\section{Introduction}

Non-Linear Sigma Models (NLSM) 
are perhaps the most interesting among two-dimensional (2D) models of
quantum field theory, both in terms of the mathematical 
physics involved, and especially in terms of their applications.
As  a classical field theory,  NLSM in the simplest setup  
describe  harmonic maps from a 2D surface 
to a Riemannian manifold (the target space). In physics NLSM were
first introduced in the beginning of the 60's as four-dimensional
phenomenological models describing the effective interaction of mesons
in the chiral limit. The interest to 2D NLSM was inspired by Polyakov
in 1975 \cite{Polyakov:1975rr} who pointed out that the $O(N)$-sigma
model (whose target space is a round $(N-1)$-sphere) is
an asymptotically free theory and, thus, could serve as an ideal laboratory for
studying the four-dimensional Yang-Mills theories.

The case of the $O(4)$-sigma model is somewhat special.
The round 3-sphere possesses a group structure so that 
the model  is equivalent to the  $SU(2)$ principal chiral field:
\bea\label{tsasopsa}
{\cal A}_{PC}=\frac{1}{4 u}\ \int\rd^2 x\
{\rm Tr}(\partial_\mu {\boldsymbol g}\,\partial^\mu
{\boldsymbol  g}^{-1})\qquad \qquad \big(\,{\boldsymbol g}\in SU(2)\,\big)\ ,
\eea
where
the  coupling constant   coincides with the inverse square of the radius  ---
the only  metric parameter of the round sphere.
In the work  \cite{Fateev:1996ea}  it was discovered  
that the following  two parameter  deformation of  the  principal chiral field
\bea\label{ospssa}
{\cal A}_{SS}=\int\rd^2 x\
\frac{u\,{\rm Tr}(\partial_\mu {\boldsymbol g}\,\partial^\mu {\boldsymbol g}^{-1})+2l\, (L_\mu^3)^2+2r\,  (R_\mu^3)^2}
{ 4 (u+r)(u+l) -r l\ \big({\rm Tr}(
{\boldsymbol g}\,\sigma_3\, {\boldsymbol g}^{-1}\,\sigma_3)\big)^2}
\eea
is a renormalizable NLSM
within a three-dimensional
space of couplings $(u,r,l)$ at the one-loop level 
(here $L_\mu^3$ and $R^3_\mu$ stands for the left and right currents:
$L_\mu^3:=\frac{1}{2\ri}\ {\rm Tr}(\partial_\mu {\boldsymbol g}\,{\boldsymbol g}^{-1}\sigma_3)$,
$R^3_\mu:=
\frac{1}{2\ri}\ {\rm Tr}(  {\boldsymbol g}^{-1}\partial_\mu  {\boldsymbol g}\sigma_3)$).
The following 
combinations  of parameters turned out to be renormalization group
(RG) invariant:
\bea\label{asisiosa}
a_1,\  a_2> 0\,:\ \ \ \ \ 
a_1\, a_2=\frac{\pi^2}{4\sqrt{(u+r)(u+l) r l}}\ ,\ \ \ \ \ 
a_1^2-a_2^2=\frac{\pi^2}{4}\ \frac{u (r-l)}{(u+r) (u+l) r l}\ .
\eea
Moreover, Fateev presented a set of convincing arguments in favor of
the quantum   integrability of the model \eqref{ospssa}.
In particular, he  argued  that  its  spectrum is  generated by two
massive doublets of the same mass whose 
2-particle $S$-matrix has the form of a direct product $(-S_{a_1}
\otimes S_{a_2})$ of two $U(1)$-symmetric solutions of the $S$-matrix
bootstrap  equations. 
For this  reason the above two-parameter deformation 
of the $O(4)$-sigma model  
was named the $SS$-model. Also, it is worth noting, that 
$S_a$ coincides with the soliton
$S$-matrix \cite{Zamolodchikov:1978xm} in the quantum sine-Gordon theory 
with the renormalized coupling constant $a$.

In this work we impose the
twisted boundary condition for the matrix valued field ${\boldsymbol g}$,
\bea\label{suaopspsa}
{\boldsymbol g}(t, x+R)=\re^{\ri\pi k_2\sigma_3}\ {\boldsymbol g}(t, x)\ 
\re^{\ri\pi k_1\sigma_3}\ .
\eea
The space of states of the theory then  splits
into sectors
characterized  by a pair
of
``winding'' numbers, ${\bf k}=(k_1,k_2)$.
The ground-state  in  each sector
is referred to below as the $k$-vacuum and the corresponding  energy 
is denoted by  $E^{\rm (vac)}_{\bf k}$.

\begin{figure}
\centering
\psfrag{a}{$ -a_1$}
\psfrag{b}{$- a_1+1$}
\psfrag{c}{$- a_1+2$}
\psfrag{d}{$-1$}
\psfrag{e}{$0$}
\psfrag{f}{$1$}
\psfrag{g}{$a_2-2$}
\psfrag{h}{$a_2-1$}
\psfrag{k}{$a_2$}
\psfrag{m}{$MR\,\cosh(\theta)$}
\includegraphics[width=14  cm]{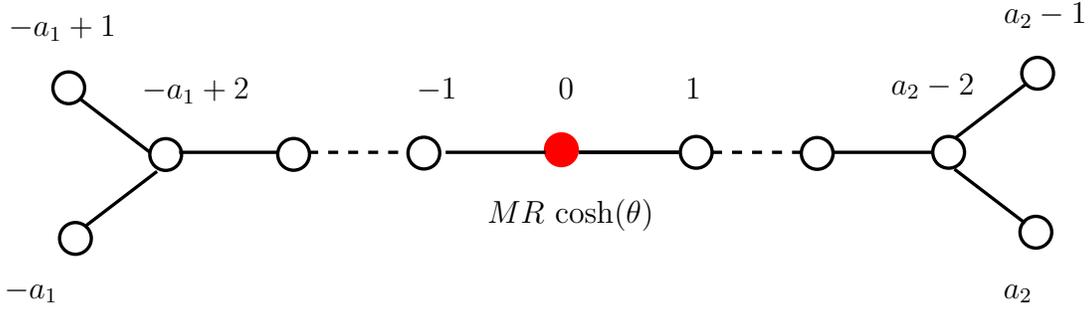}
\caption{Incidence diagram for the TBA system describing  the vacuum
  energy at the sector $k_1=k_2=0$ in the case $a_{1},
  a_{2}=2,\,3,\,4\ldots\ .$ The source term is indicated near the
  corresponding node.} 
\label{fig1}
\end{figure}

The lowest vacuum energy $E^{({\rm vac})}_{{\bf k}=0}$, can be
calculated in the framework of the Thermodynamic Bethe Ansatz (TBA)
approach. For the simplest case of integer parameters $a_1, 
a_2=2,\,3,4,\ldots$, the required TBA equations were obtained in
\cite{Fateev:1996ea}.  These equations are encoded by the incidence
diagram shown in Fig.\ref{fig1}, which has one massive node.\footnote{%
As noted in \cite{Fendley:1991ve}, 
if a model has an $S$-matrix in 
the form of a direct product $(-S_G\otimes S_H)$ and the TBA equations
for the models described by $S$-matrices $S_G$ and
$S_H$ are encoded by Dynkin-like
diagrams of type $G$ and $H$, each having one massive node, then the
TBA equations for the model with the direct product $S$-matrix are
obtained by ``gluing'' together the individual TBA equations  at their
massive nodes.  This prescription, when applied to the
$SS$-model with integer $a_1,a_2\ge2$, leads to a TBA system
whose incidence diagram is shown in Fig.\ref{fig1}.}
Subsequently, in Ref.\cite{Hegedus:2003xd},
these equations were generalized 
to a system of Non-Linear Integral Equations (NLIE) 
\cite{Klummpe:1991vs,Destri:1994bv}
which allows one to calculate
$E^{({\rm vac})}_{{\bf k}=0}$
for any values  of  $a_1,a_2\geq 2$. Moreover, 
the  ${\bf k}=0$ case of the 
undeformed $O(4)$-sigma model was separately considered in
Refs.\cite{Balog:2003yr,Gromov:2008gj, Caetano:2010zd}.
However, to the best of our knowledge, the problem of calculating the
$k$-vacuum energies for general values of $a_i$ and $k_i$ is beyond
the scope of traditional approaches of integrable quantum field
theory.
The purpose of this note is to extend a result of earlier work
\cite{Lukyanov:2013wra} and conjecture an exact formula
for the $k$-vacuum energy in the $SS$-model in the   general  case.

\section{UV/IR  behavior of  $k$-vacuum energy}

Although $E^{({\rm vac})}_{{\bf k}}$ is a rather complicated function
of the parameters, its leading small-$R$ (i.e., UV) and large-$R$ (IR)
behavior can be obtained via a simple and intuitive analysis which is
based on the dual form of the $SS$-model proposed in
Ref.\cite{Fateev:1996ea}.

The dual description is formulated in terms of  three Bose fields
governed by the Toda-like  Lagrangian
\bea\label{aposoasio}
{\widetilde {\cal L}}_{SS}= 
\frac{1}{16\pi}\ \sum_{i=1}^3  (\partial_\mu\varphi_i \partial^\mu\varphi_i)^2
+2\mu\
\big(\, \re^{b\varphi_3}\ \cos(\alpha_1\varphi_1+\alpha_2\varphi_2)+\re^{-b\varphi_3}\
\cos(\alpha_1\varphi_1-\alpha_2\varphi_2)\,\big)\ ,
\eea
where
\bea\label{isoio}
\alpha_i=\frac{1}{2}\ \sqrt{a_i}\ ,\ \ \ \ \ \ \ \ \ \ \ \ b=\frac{1}{2}\ \sqrt{a_1+a_2-2}\
\eea
and the dimensionfull coupling $\mu$  is related to the soliton mass as
\bea\label{saopsao}
M=2\mu\ \frac{\Gamma(2\alpha_1^2)\Gamma(2\alpha_2^2)}{\Gamma(2 \alpha_1^2+2\alpha_2^2)}\ .
\eea
The soliton charges
$q_i=0,\pm 1,\pm 2\ldots$,
corresponding to  the  factors $S_{a_i}\ (i=1,2)$ in the direct 
product $(-S_{a_1}\otimes S_{a_2})$,
appear  through the quasiperiodic boundary conditions
imposed on the  dual fields:
\bea\label{saopssap}
\varphi_1(x_1+R)=\varphi_1(x_1)+\frac{\pi}{\alpha_1}\ (q_2+q_1)\ ,\ \ \ \
\varphi_2(x_1+R)=\varphi_2(x_1)+\frac{\pi}{\alpha_2}\ (q_2-q_1)\ .
\eea
In their turn,
the winding numbers are interpreted as quasimomenta.
Due to the periodicity of 
the potential terms in $\varphi_j\  (j=1,2)$, the stationary states
can be chosen to  be the Floquet  states characterized by the pair  ${\bf k}=(k_1,k_2)$:
\bea\label{sapsapo}
\varphi_i\mapsto\varphi_i+2\pi/\alpha_i\ :\
\ \ |\,\Psi_{\bf k}\,\rangle\mapsto\re^{2\pi\ri k_i}\  |\,\Psi_{\bf k}\,\rangle\ .
\eea

The form of the dual Lagrangian suggests that for 
small $R$
\bea\label{saopasasps}
E_{\bf k}^{({\rm vac})}\approx \frac{\pi}{R}\ \left(\,-\frac{1}{2}+
\frac{p^2_0}{4b^2}+ a_1 k_1^2+a_2 k_2^2\,\right)\ .
\eea
Since values of the field $\varphi_3$
is effectively  restricted within the segment of length $\big(- 2b\,\log(\mu R)\big)$, 
the corresponding ``zero-mode momentum''  $p_0$
is not arbitrary. It  is determined through a certain quantization condition, similar 
to that discussed in Ref.\cite{Zamolodchikov:1995aa} in the context of
the quantum sinh-Gordon model.
Assuming that
\bea\label{asopssoap}
|a_1k_1\pm a_2 k_2|<1\ ,
\eea
the original consideration from \cite{Zamolodchikov:1995aa}
can be applied to  the $SS$-model  yielding
\bea\label{ospsasaop}
-  \frac{p_0}{b^2} \ \log\Big(\frac{\mu R}{8b^2}\Big)+  \delta^{({\bf q}=0)}(p_0)\approx  2\pi \ ,
\eea
with
\bea\label{asopssap}
\delta^{({\bf q})}(p)=-\ri\
\log\Big(S^{(q_1)}(p|a_1k_1-a_2 k_2)\, S^{(q_2)}(p|a_1k_1+a_2 k_2)\Big)
\ \ \ \ \ \ \ \ \ \ \big(\,\delta^{({\bf q})}(0)=0\,\big)\ .
\eea
Here $S^{({\bf q})}(p|\lambda)$ stands for the so-called 
``reflection amplitude'' for the sine-Liouville model \cite{ZAM}
\bea\label{ospssap}
S^{({ q})}(p|\lambda)=
\frac{\Gamma(\frac{1+|q|}{2}+
\frac{\lambda}{2}-\frac{\ri p}{2})
\Gamma(\frac{1+|q|}{2}-\frac{\lambda}{2}-\frac{\ri p}{2})}
{\Gamma(\frac{1+|q|}{2}+\frac{\lambda}{2}+\frac{\ri p}{2})
\Gamma(\frac{1+|q|}{2}-\frac{\lambda}{2}+\frac{\ri p}{2})}\
\frac{\Gamma(1+\ri p)\Gamma(1+\frac{\ri p}{4b^2})}
{\Gamma(1-\ri p)\Gamma(1-\frac{\ri p}{4b^2})}\ .
\eea

In the IR  limit the $k$-vacuum energy is composed of an extensive part proportional
to the length of the system
\bea\label{sososasa}
E_{\bf k}^{({\rm vac })}=R\,{\cal E}_0+o(1)\ \ \ \ \ \ \ \ \ {\rm as}\ \ \ \ \ \ R\to\infty\ .
\eea
The exact form of the specific bulk energy was found in \cite{Fateev:1996ea}.
It is  expressed 
through the soliton mass $M$ as
\bea\label{sopsosap}
{\cal E}_0=-\frac{M^2}{4}\ \frac{\sin(\frac{\pi }{2} a_1)
\sin(\frac{\pi }{2} a_2)}
{\sin(\frac{\pi}{2} (a_1+a_2))}\ .
\eea
In the case $a_1,\,a_2>1$, when the fundamental particles do not form bound states,
the leading correction to \eqref{sososasa}
comes  from  virtual soliton and 
antisoliton trajectories winding once around the space circle.
These trajectories should be counted with the  phase factor  
$\re^{\ri\pi(\sigma_1 k_1+\sigma_2k_2)}$, where $\sigma_{1,2}=\pm 1$. 
Therefore, summing over the four possible sign combinations one obtains
\bea\label{saosa}
E_{\bf k}^{({\rm vac})}=R\,{\cal E}_0
-{\textstyle \frac{4}{\pi}}\, \cos(\pi k_1)\, \cos(\pi k_2)\
M\, K_{1}(MR)+\big({\rm multiparticle}\big)\ \ \ \ \ (a_{1,2}>1)
\eea
(here
$K_1(r)$ stands for  the conventional Bessel function).
Note that similar arguments were  originally applied to the quantum
sine-Gordon model by Al. Zamolodchikov in Ref.\cite{Zamolodchikov:1994uw}.

\begin{figure}
\centering
\includegraphics[width=8  cm]{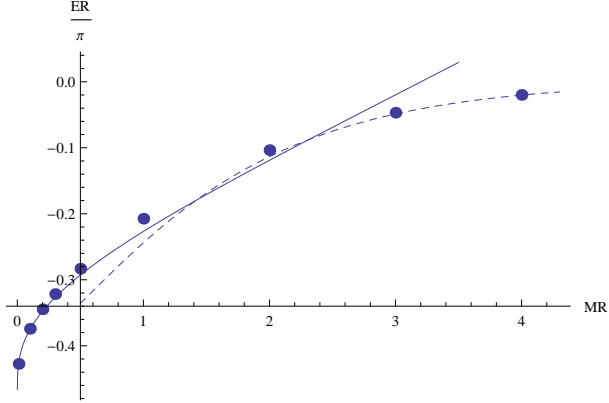}
\caption{Numerical values of the dimensionless $k$-vacuum energy 
$\frac{R}{\pi}\ E^{({\rm vac })}_{\bf k}$ versus $MR$ for $a_1=2,\, a_2=3$
and
$k_1=k_2=0$. The solid   and  dashed  lines follow from UV \eqref{saopasasps},
and IR \eqref{saosa} asymptotic formulas, respectively.
The heavy dots were obtained
by means of a numerical solution of the TBA system encoded by  the
incidence diagram of 
Fig.\ref{fig1}.}
\label{fig2}
\end{figure}

In Fig.~\ref{fig2} the UV/IR asymptotic formulae are compared 
with the results of a numerical solution of the TBA system  described by the 
incidence diagram from Fig.\ref{fig1}.

\section{Exact $k$-vacuum energy}

\subsection{Fateev model}

The model governed by the  Lagrangian
\bea\label{asoasio}
{ {\cal L}}_{F}= \frac{1}{16\pi}\ \sum_{i=1}^3  (\partial_\mu\varphi_i \partial^\mu\varphi_i)^2
+2\mu\
\big(\, \re^{\ri \alpha_3\varphi_3}\ \cos(\alpha_1\varphi_1+\alpha_2\varphi_2)+\re^{-\ri \alpha_3\varphi_3}\
\cos(\alpha_1\varphi_1-\alpha_2\varphi_2)\,\big)\ ,
\eea
where the coupling constants $\alpha_i$ are subjected to a single constraint
\bea\label{asopssa}
\alpha_1^2+\alpha_2^2+\alpha_3^2=\frac{1}{2}\ ,
\eea
will be referred to below as the {\it Fateev model}.  In the case when
$\alpha_1,\ \alpha_2$ are real while $\alpha_3$ is pure imaginary
({\it unitary regime}), the Lagrangian \eqref{asoasio} is real and
coincides with the dual Lagrangian ${\widetilde {\cal L}}_{SS}$ provided
$\alpha_3=-\ri b$.  In the {\it symmetric regime} all the coupling
constant $\alpha_i$ are real, the Lagrangian \eqref{asoasio} is
completely symmetric under simultaneous permutations of the real
fields $\varphi_i$ and couplings $\alpha_i$. Despite that the theory
is apparently non-unitary in this case, one can still address the
problem of calculation of the $k$-vacuum energies.  Since the
Lagrangian ${\cal L}_F$ in the symmetric regime is invariant
under the transformations $\varphi_i\mapsto \varphi_i+2\pi\alpha_i$ with
$i=1,2,3$, the $k$-vacuum energies are labeled by the triple of quasimomenta
${\bf  k}=(k_1,k_2,k_3)$ (contrary to the unitary regime where 
${\bf  k}=(k_1,k_2)$).  The short distance expansion of 
$E_{\bf k}^{({\rm vac})}$ in the symmetric regime is considerably
simpler than in the unitary one.  
Its general structure follows from the fact that
the potential term of ${\cal L}_F$ with $\alpha_i>0$ is a uniformly
bounded perturbation for any values of the dimensionless parameter
$\mu R$. Therefore the conformal perturbation theory can be applied
literally yielding an expansion
\bea\label{osapsoa}
{\rm Symmetric\ regime}\ :\ \ \ \ \frac{R}{\pi}\ E_{\bf k}^{({\rm vac})}=-\frac{1}{2}+
\sum_{i=1}^3 (2\alpha_i k_i)^2-\sum_{n=1}^\infty e_n\ (\mu R)^{4n}\ .
\eea
An exact  formula for 
the $k$-vacuum energies in the symmetric regime was proposed in Ref.\cite{Lukyanov:2013wra}.
Bellow   we argue that essentially the same formula
actually holds in both regimes of the
Fateev model.

\subsection{Regular solutions of the shG equation}

\begin{figure} 
\centering
\psfrag{a}{$\pi a_1$}
\psfrag{b}{$\pi a_2$}
\psfrag{c}{$\frac{\pi a_1}{2}$}
\psfrag{d}{$\frac{\pi a_2}{2}$}
\psfrag{e}{$\frac{\pi a_3}{2}$}
\psfrag{w1}{$w_1$}
\psfrag{w2}{$w_2$}
\psfrag{w3}{$w_3$}
\psfrag{bw3}{${\bar w}_3$}
\includegraphics[width=5  cm]{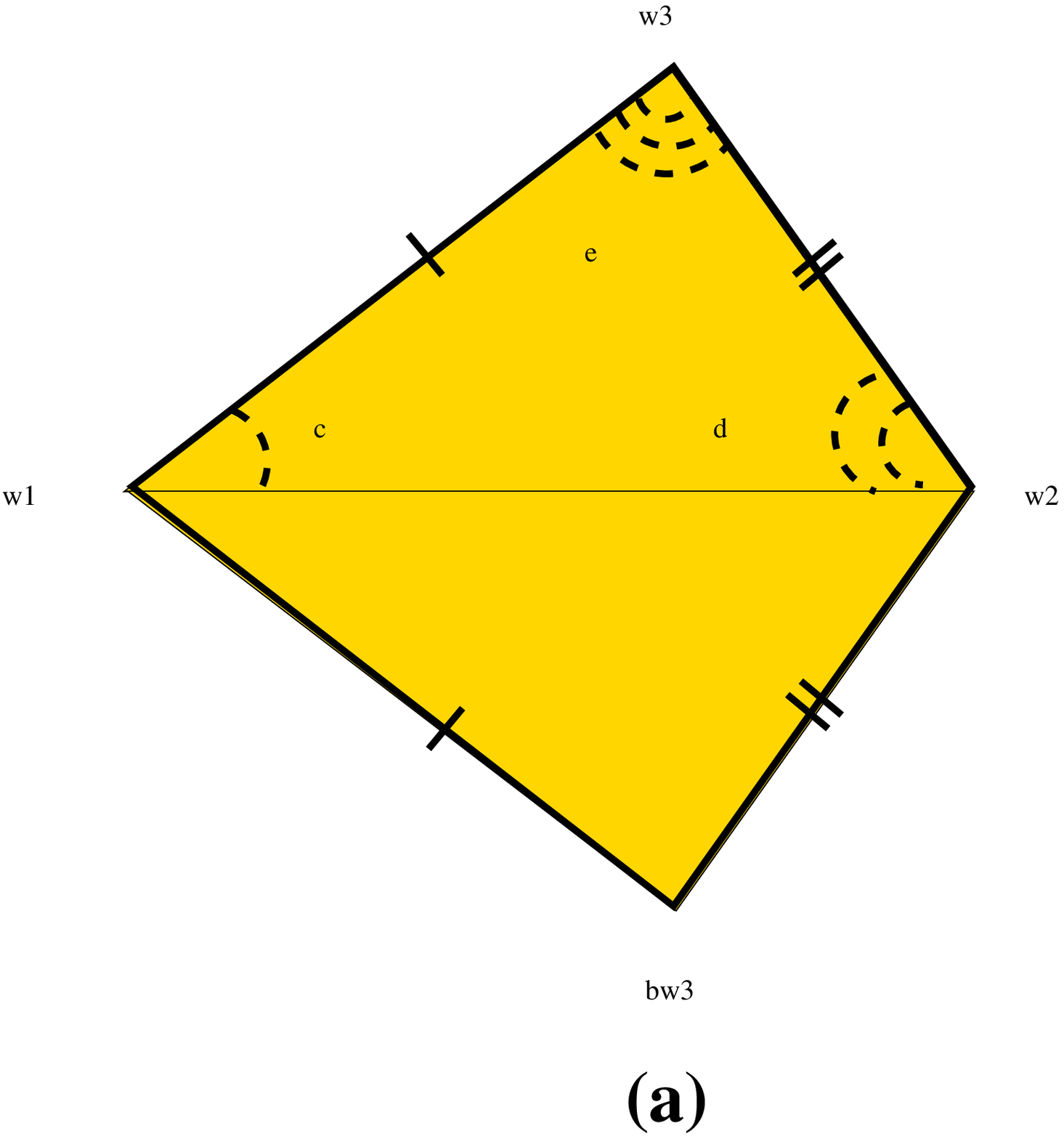}\ \ \ \ \ \ \ \ \ \ \ \ \ \ \ \ \ 
\includegraphics[width=4  cm]{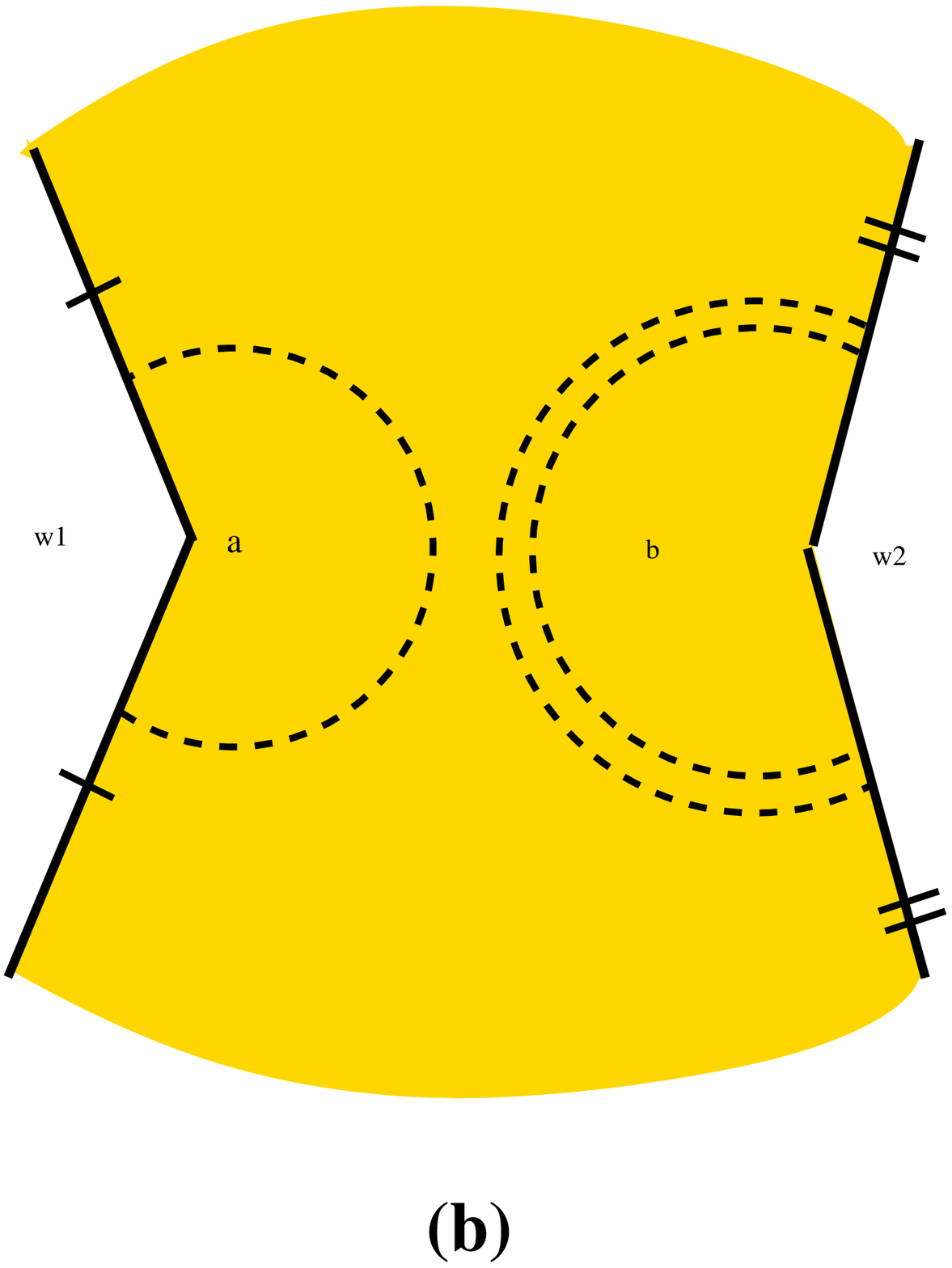}
\caption{The image of the thrice-punctured sphere in the complex $w$-plain: (a)
  for the case $a_{1,2,3}>0$ (regime I); (b) for the case $a_{1,2}>0$ and
  $a_3<0$ (regime II).}
\label{fig3}
\end{figure}

Consider the  classical partial differential equation
\bea\label{asospsosap}
\partial_z\partial_{\bar z}{\hat \eta}-\rho^2\ |{\cal P}(z)|\, \big(\re^{2{\hat \eta}}-
 \re^{-2{\hat \eta}}\big)=0
\eea
where
\bea\label{oasioaq}
{\cal P}(z)=\frac{(z_3-z_2)^{a_1}\,(z_1-z_3)^{a_2}\,(z_2-z_1)^{a_3}}
{(z-z_1)^{2-a_1}(z-z_2)^{2-a_2}(z-z_3)^{2-a_3}}\ .
\eea
and $\bar z$ denotes the complex conjugate of $z$. 
Here $\rho$ is a real parameter and $a_i$ $(i=1,2,3)$ are also real
and satisfy the condition
\bea\label{iasoaisa}
a_1+a_2+a_3=2\ .
\eea
The variable $z$ is regarded as a complex coordinate on
${\mathbb C}{\mathbb P}^1\backslash\{z_1,z_2,z_3\}$, 
the Riemann sphere with three punctures.
Due to the  relation \eqref{iasoaisa}, 
${\cal P}(z)(\rd z)^2$  is
a quadratic differential under $\mathbb{PSL}(2,\mathbb{C})$
transformations, so that   the punctures
can be sent  to  any prescribed positions, say
$(z_1,z_2,z_3)=(0,1,\infty)$.
Then the change of variables
\bea\label{opsasoap}
w=
\rho\ \int \rd z\ z^{\frac{a_1}{2}-1}\, (1-z)^{\frac{a_2}{2}-1}
\eea
brings \eqref{asospsosap} to the standard form of  the sinh-Gordon
(shG) equation, 
\bea\label{asososap}
\partial_w\partial_{\bar w}{\hat \eta}-\re^{2{\hat \eta}}+
\re^{-2{\hat \eta}}=0\ .
\eea
In the case when $a_1,a_2,a_3$ are all positive Eq.\eqref{opsasoap}
defines the Schwarz-Christoffel mapping, transforming the upper and
lower half-planes correspondingly
to the triangles $(w_1,w_2,w_3)$ and
$(w_1,w_2,{\bar w}_3)$, depicted in Fig.\,\ref{fig3}a.
Note, that the adjacent sides of the resulting polygon
$(w_1$,\ $w_2$,\ $w_3$,\ ${\bar w}_3)$ should be
identified to form a topological 2-sphere. 
In the case when $a_3<0$, but $a_{1}, a_{2}>0$,  
the image of the  punctured sphere 
is shown   in   Fig.\,\ref{fig3}b. Again,
the adjacent  rays should be properly identified.
In this way Eq.\eqref{asospsosap} 
on the thrice-punctured sphere
can be equivalently formulated as the shG equation in
the domains shown in  Fig.\,\ref{fig3}a and Fig.\,\ref{fig3}b,
corresponding to the two cases
\bea\label{asoposapsa}
&{\rm Regime\  I\ }\ \ \ \ &:\ \ \ a_1>0,\  a_2>0\ ,\ \ \ a_3=2-a_1-a_2>0\\
&{\rm Regime\  II}\,\ \ \ &:\ \ \ a_1>0,\  a_2>0\ ,\ \ \ a_3=2-a_1-a_2<0\ 
.\nonumber 
\eea

We will consider {\it regular} solutions to \eqref{asososap}, defined
by the following two requirements.  First, the regular solution should
be a {\it smooth}, {\it single valued}, {\it real\/} function on the
punctured sphere ${\mathbb C}{\mathbb P}^1\backslash\{z_1,z_2,z_3\}$
or, equivalently (when the complex coordinate $w$ is employed) in the
domains shown in Fig.\,\ref{fig3} with properly identified edges.
Second, the regular solution must develop the proper asymptotic
behavior in the vicinity of the punctures. For regime I there is 
the freedom to control the asymptotic behavior of ${\hat \eta}$ at
each of the three punctures, or, equivalently, at each vertex $w_i$ in
Fig.\ref{fig3}a. Namely,
\bea\label{roapsoa}
{\hat \eta}\to 2\,l_i\ \log|w-w_i|+O(1)\,,\ \ \  \mbox{when}\ \ w\to w_i\,,
\eea
where
\bea\label{saoposa}
-\frac{1}{2}<l_i\leq 0
\eea
denote free parameters\footnote{%
For $l_i=-\frac{1}{2}$ the leading asymptotics \eqref{roapsoa} should be
replaced by
$${\hat \eta}\to -\log\Big(|w-w_i|\log\big({\textstyle
  \frac{4}{|w-w_i|}}\big)\Big)+O(1)\ .$$}.
For regime II, when $a_3<0$, the third  puncture is mapped to
the infinity of the domain, shown in Fig.\ref{fig3}b, 
and we require that
\bea\label{osapssao}
{\rm Regime\  II}\ \ \ :\ 
\ \ \ \ \ \ \ {\hat \eta}\to 0\ \ \ \ \ \ {\rm as}\ \ \ \ |w|\to \infty\ ,
\eea
whereas the asymptotic behavior in the vicinity of $w=w_1,\,w_2$,\   
is still described by \eqref{roapsoa} 
with two free parameters \eqref{saoposa}. It turns out that the
solution of the shG equation, satisfying the above regularity conditions,
exists and is unique for both regimes I and II.

\subsection{Main conjecture}

Define the functional 
\bea\label{soapsopsa}
{\mathfrak F}(\rho)=-{ \frac{8}{\pi}}\  \int\rd^2 w\  \sinh^2({\hat\eta})+
\sum_{i}  a_il_i^2\ ,
\eea
where ${\hat \eta}$ is a  regular 
solution and the summation index $i$ takes the values $i=1,2,3$ and $i=1,2$
for the regimes I and  II,  respectively.
The additive  constant in \eqref{soapsopsa} is chosen 
to provide the normalization condition
\bea\label{sopssapos}
\lim_{\rho\to\infty}{\mathfrak F}(\rho)=0\ .
\eea
Now we can extend the conjecture of Ref.\cite{Lukyanov:2013wra}
and propose the expression for the $k$-vacuum energies, which is valid
for both considered regimes,
\bea\label{ospsopsa}
\frac{R}{\pi}\  E_{\bf k}^{({\rm vac})}={\mathfrak F}(\rho)-4\rho^2  \prod_{i=1}^3\gamma\big({\textstyle\frac{a_i}{2}}\big)\ ,
\eea
where $\gamma(x):=\frac{\Gamma(x)}{\Gamma(1-x)}$.
This formula should be supplemented with the relations between the
parameters of   
quantum and  classical problems:
\bea\label{soassap}
\mu R=2\,\rho\ ,\ \ \ \ \ \ \alpha_i^2=\frac{a_i}{4}\ ,\ \ \ \ |k_i|=l_i+\frac{1}{2}\ .
\eea

In  the case of the symmetric regime, formula \eqref{ospsopsa} 
can be checked, in principle, perturbatively.
Namely, let us return to
the original variable $z$ and
replace  ${\hat\eta}$ by $\eta={\hat \eta}+\frac{1}{2}\ \log|\rho^2{\cal P}|$.
This  brings  \eqref{asospsosap} to the form of the modified shG equation:
\bea\label{asospsosapr}
\partial_z\partial_{\bar z}{ \eta}-\re^{2{ \eta}}+
\rho^4\ |{\cal P}|^2\  \re^{-2{ \eta}}=0\ .
\eea
For the regular solution the third term in \eqref{asospsosapr} can be treated perturbatively
even in the nearest neighbor of each 
puncture and  the RHS of \eqref{ospsopsa}
admits a Taylor  expansion (see \cite{Lukyanov:2013wra} for details):
\bea\label{sopssa}
{\rm Regime\ I}\ \ :\ \ \ \ {\mathfrak F}-4\rho^2  \prod_{i=1}^3\gamma\big({\textstyle\frac{a_i}{2}}\big)=
\sum_{n=0}^\infty {\mathfrak f}_n\ \rho^{4n}\ .
\eea
On the other hand,
the LHS of \eqref{ospsopsa}
possesses a series expansion
\eqref{osapsoa} which,  in principle, can be obtained using
the conformal perturbation theory. Thus, in the symmetric regime (regime I), 
both sides of
\eqref{ospsopsa} can be understood perturbatively and 
the conjectured  relation
implies that
the corresponding expansion coefficients are simply related:
${\mathfrak f}_n=-2^{4n}\ e_n$.

\begin{figure}
\centering
\includegraphics[width=8  cm]{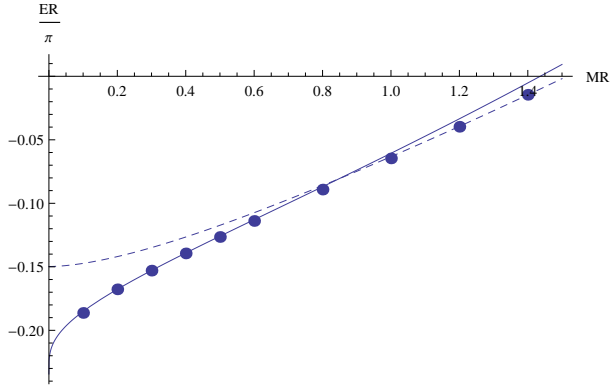}
\caption{Numerical values of the dimensionless 
$k$-vacuum energy $\frac{R}{\pi}\ E^{({\rm vac})}_{\bf k}$ versus the
  variable $r=MR$ for $a_1=1.7,\, a_2=1.5,\,
k_1=\frac{4}{17}=0.235\ldots,\ k_2=\frac{1}{3}$. The solid   and  dashed  
lines represent the small-$R$, \eqref{saopasasps},
and large-$R$ asymptotics \eqref{saosa}, respectively.
The heavy dots represent  the LHS
of  \eqref{ospsopsa} calculated from numerical 
solutions of the shG equation.  The corresponding numerical values are
presented in Table\,\ref{tab1}.}
\label{fig5}
\end{figure}

The situation is somewhat different in the unitary regime (regime II).
Of course, the RHS of \eqref{ospsopsa} in this regime
is still well defined.
However,  the conformal perturbation theory cannot be applied
literally in this case. More generally, at the moment, it is not
entirely clear how one can calculate the LHS of \eqref{ospsopsa}
for arbitrary values of $a_i$ and $k_i$ in the $SS$-model. 
In particular, as was mentioned earlier, the knowledge of the exact
$S$-matrix is not of much help in solving this problem.
Therefore, as a first step in proving the correspondence
\eqref{ospsopsa}, it would be desirable to {\it derive} the UV
and IR asymptotics of $E_{\bf k}^{(\rm vac)}$, discussed above, from the
differential equation side.   
Fortunately, this could be done analytically by using an
auxiliary linear problem associated with the shG equation \eqref{asososap}.
The derivation is rather technical and will be published elsewhere. 
Here we only present the results of our numerical work in support of the
conjecture \eqref{ospsopsa}. The shG equation has been solved numerically 
for various sets of the parameters $a_i$ and $k_i$. We found that the
resulting   
values of the RHS of \eqref{ospsopsa} are in good agreement with 
the UV and IR asymptotic formulae \eqref{saopasasps} and \eqref{saosa}. 
Some (small) part  of the available numerical data is presented 
in Fig.\,\ref{fig5} and Table\,\ref{tab1}.

\begin{table}
[!ht]
\begin{center}
\begin{tabular}{| c || l | l | l|}
\hline \rule{0mm}{3.6mm}
$MR$ & r.h.s of Eq.\eqref{ospsopsa} & Eq.\eqref{saopasasps} & Eq.\eqref{saosa}\\
\hline
$0.1$  &\ \ \  \ \    -0.18631 &  -0.18510 &-0.14729    \\ 
$0.2$  &\ \ \  \ \    -0.16773 &  -0.16770 &-0.14197    \\
$0.3$  &\ \ \  \ \    -0.15230 &  -0.15288 &-0.13487    \\
$0.4$  &\ \ \  \ \    -0.13945 &  -0.13913 &-0.12654    \\
$0.5$  &\ \ \  \ \    -0.12651 &  -0.12589 &-0.11731    \\
$0.6$  &\ \ \  \ \    -0.11390 &  -0.11288 &-0.10739    \\
$0.8$  &\ \ \  \ \    -0.08919 &  -0.08695 &-0.08605    \\   
$1.0$  &\ \ \  \ \    -0.06463 &  -0.06061 &-0.06327    \\   
$1.2$  &\ \ \  \ \    -0.03973 &  -0.03346 &-0.03941    \\
$1.4$  &\ \ \  \ \    -0.01456 &  -0.00524 &-0.01461    \\
\hline
\end{tabular}
\end{center}
\caption{The dimensionless $k$-vacuum energy  $\frac{R}{\pi}\ E_{\bf
    k}^{({\rm vac})}$ as a function of the variable $MR$.}
\label{tab1}
\end{table}

\section{Conclusion}

In this work we propose the formula \eqref{ospsopsa}
for the $k$-vacuum energy in the 
$SS$-model for arbitrary values of parameters $a_1$ and $a_2$.  
Our conjecture is in agreement with expected
UV and IR asymptotics of $E^{({\rm vac)}}_{\bf k}$. This can be 
shown analytically  by using
the auxiliary linear problem associated with the shG equation.
The derivation exploits standard
techniques of the inverse scattering method and we did not present 
it in this short note.
Instead, we reported a few numerical verifications
supporting  our conjecture.
Another analytical result, which also remained beyond  the scope of the  paper, is
a system of integral equations for the calculation of  $E^{({\rm
    vac)}}_{\bf k}$. 
We are planing  to discuss it 
in a separate publication.

Finally, note that our proposal 
for the $k$-vacuum energy can be extended to 
the whole energy spectrum of the $SS$-model, 
following the approach of Ref.\cite{Bazhanov:2013cua} previously applied to
the symmetric regime of the Fateev model.

As for future work,
we believe that the most  intriguing generalization
lies in study of a remarkable two-parameter deformation
of the general principal chiral field discovered by
Klim$\check{\rm c}$ik \cite{Klimcik:2008eq,Klimcik:2014bta}.
In Ref.\cite{Hoare:2014pna} 
it was   pointed out that in the simplest
$SU(2)$ case the Klim$\check{\rm c}$ik NLSM coincides with the
$SS$-model.

\section*{Acknowledgments}

The authors are grateful to A.~B. Zamolodchikov for fruitful discussions.

\bigskip
\noindent
Part of this work was done during the visit of the  third author to IPhT at
CEA Saclay in June 2014. SL would like to express his sincere gratitude
to members of the laboratory and especially Didina  Serban and Ivan Kostov
for their kind hospitality and interesting discussions. 
The research of VB and GK was
partially supported by the Australian Research Council.

\end{document}